\documentclass{elsart}
\usepackage{amssymb}

\begin{document}
\begin{frontmatter}
\title{DETERMINANTS OF BLOCK TRIDIAGONAL MATRICES}
\author{Luca Guido Molinari}
\address{Dipartimento di Fisica, Universit\`a degli Studi di Milano,\\
and INFN, Sezione di Milano, Via Celoria 16, Milano, Italy}
\ead{luca.molinari@mi.infn.it}
\begin{abstract}
An identity is proven that evaluates the determinant
of a block tridiagonal matrix with (or without) corners as the determinant of
the associated transfer matrix (or a submatrix of it).
\end{abstract}
\end{frontmatter}

\section{Introduction}
A tridiagonal matrix with entries given by square matrices is a 
block tridiagonal matrix; the matrix is banded if off-diagonal
blocks are upper or lower triangular. Such matrices are of great 
importance in numerical analysis and physics, and 
to obtain general properties is of great utility.  
The blocks of the inverse matrix of a block tridiagonal 
matrix can be factored
in terms of two sets of matrices\cite{Meurant92}, and decay rates of 
their matrix elements have been investigated\cite{Nabben99}. 
While the spectral properties of tridiagonal matrices have been
under study for a long time, those of tridiagonal block matrices are 
at a very initial stage\cite{Aptekarev84,Korotyaev06}.  

What about determinants?
A paper by El-Mikkawy\cite{ElM} on determinants of
tridiagonal matrices triggered two interesting generalizations for
the evaluation of determinants of block-tridiagonal and general complex 
block matrices, respectively by Salkuyeh\cite{Salkuyeh06} and 
Sogabe\cite{Sogabe07}. 
These results encouraged me to re-examine
a nice identity that I derived in the context of transport\cite{Molinari97}, 
and extend it as a mathematical result 
for general block-tridiagonal complex matrices.

For ordinary tridiagonal matrices, determinants can be evaluated 
via multiplication of $2\times 2$ matrices: 
\begin{eqnarray}
&&\det\left[\begin{array}{cccc} 
  a_1     & b_1          & {} & c_0  \\
  c_1     & \ddots    & \ddots       &        {}       \\
 {}       &  \ddots    & \ddots  & b_{n-1}     \\
 b_n   &     {}       & c_{n-1}   &  a_n  
\end{array}\right] =
(-1)^{n+1}(b_n\cdots b_1 + c_{n-1}\cdots c_0)\nonumber \\
&&\qquad +\,{\rm tr}\left[
\left(\begin{array}{cc}a_n & -b_{n-1} c_{n-1} \\1 & 0 \end{array}\right)
\cdots \left(\begin{array}{cc} a_2 & -b_1c_1 \\1 & 0 \end{array}\right)
\left(\begin{array}{cc} a_1 & -b_nc_0 \\1 & 0 \end{array}\right)
\right ]\label{trid0}
\end{eqnarray}
\begin{equation}
\label{trid}
 \det \left[\begin{array}{cccc} 
  a_1     & b_1       & {}        & {}     \\
  c_1     & \ddots    & \ddots    & {}     \\
  {}      & \ddots    & \ddots    & b_{n-1}\\
  {}      &  {}       & c_{n-1}   &  a_n  
\end{array}\right] \,=\, \left [ 
\left(\begin{array}{cc}a_n & -b_{n-1} c_{n-1} \\1 & 0 \end{array}\right)
\cdots
\left(\begin{array}{cc} a_2 & -b_1 c_1 \\1 & 0 \end{array}\right) 
\left(\begin{array}{cc} a_1 & 0 \\1 & 0 \end{array}\right)
\right]_{11} 
\end{equation}
Do these procedures generalize to block-tridiagonal matrices? 
The answer is affirmative. If the matrix has corner blocks, the 
determinant is proportional to that of an associated transfer 
matrix, in general of much smaller size. The proof is simple and is given in 
section 2. A simple modification yields a formula for the determinant when 
corner blocks are absent, and is given in section 3. The relation with 
Salkuyeh's recursion formula is then shown.

\section{The Duality Relation}
Consider the following block-tridiagonal matrix $\mathrm{M}(z)$ with blocks 
$A_i$, $B_i$ and $C_{i-1}$ ($i=1,\ldots ,n$) that are complex $m\times m$  
matrices. It is very useful to introduce also a complex parameter $z$ in the
corner blocks:
\begin{equation}
\label{mz}
 \mathrm{M}(z) =\left[\begin{array}{cccc} 
  A_1     & B_1    & {}           & \frac{1}{z} C_0  \\
  C_1     & \ddots & \ddots       &        {}        \\
  {}      & \ddots & \ddots       &      B_{n-1}     \\
 z B_n    & {}     & C_{n-1}      &         A_n  
\end{array}\right]
\end{equation}
It is required that off-diagonal blocks are nonsingular: 
$\det B_i\neq 0$ and $\det C_{i-1}\neq 0$ for all $i$. 
As it will be explained, the matrix is naturally associated with a 
{\em transfer matrix}, built as the product of $n$ matrices of size 
$2m\times 2m$:
\begin{equation}
\label{transfer}
\mathrm{T}=\left[\begin{array}{cc}
-B_n^{-1}A_n & -B_n^{-1}C_{n-1} \\I_m & 0 \end{array}\right]
\ldots
\left[\begin{array}{cc} -B_1^{-1}A_1 & -B_1^{-1}C_0 \\I_m & 0 \end{array}
\right] 
\end{equation}
where $I_m$ is the $m\times m$ unit matrix.
The transfer matrix is nonsingular, since
\begin{equation}
\label{detT}
\det \mathrm{T} = \prod_{i=1}^n \det [B_i^{-1}C_{i-1}] 
\end{equation}
The main result, the duality relation, relies on the following lemma:

\begin{lem}
\begin{displaymath}
\det\mathrm{M}(z)\, = \, \frac{(-1)^{nm}}{(-z)^m} 
\,\det[\,\mathrm{T}-z\,I_{2m}] \,
\det[\,B_1\ldots B_n\,]
\end{displaymath}
\end{lem}
Proof: The equation $\mathrm{M}(z)\Psi =0$ has a nontrivial 
solution provided that $\det\mathrm{M}(z)=0$, and corresponds to the
following linear system in terms of the blocks of the matrix and the 
components $\psi_k \in \mathbb{C}^m$ of the null vector $\Psi $: 
\begin{eqnarray}
&& A_1\,\psi_1 +B_1\,\psi_2 +z^{-1}C_0\, \psi_n\,=\,0\label{ls1} \\
&&B_k {\psi}_{k+1}+ A_k \psi_k + C_{k-1}\psi_{k-1}= 0 
\qquad (k=2,\ldots ,n-1 )\label{ls2}\\
&& z\,B_n \psi_1 +A_n\,\psi_n + C_{n-1}\,\psi_{n-1}\, =\,0 
\label{ls3}
\end{eqnarray}
The equations (\ref{ls2}) are recursive and can be put in the form 
\begin{displaymath}
\left[\begin{array}{c}\psi_{k+1}\\ \psi_k\end{array}\right] \,=\, 
\left[\begin{array}{cc} -B_k^{-1}A_k & -B_k^{-1}C_{k-1}\\ I_m & 0
\end{array}\right]
\left[\begin{array}{c}\psi_k\\ \psi_{k-1}\end{array}\right]
\end{displaymath}
and iterated. Inclusion of the boundary equations (\ref{ls1}) and (\ref{ls3}) 
produces an eigenvalue equation for the full transfer matrix (\ref{transfer})
that involves only the end vector-components:
\begin{equation}
\label{eigT}
 \mathrm {T} 
\left[\begin{array}{c} \psi_1\\  \frac{1}{z}\psi_n\end{array}\right] 
\, =\, z \,\left[\begin{array}{c} \psi_1\\ 
\frac{1}{z}\psi_n \end{array}\right]
\end{equation}
Equation (\ref{eigT}) has a nontrivial solution if and only if 
$\det [\mathrm{T}-zI_{2m}]=0$, 
which is dual to the condition $\det {\rm M}(z)=0$. 
Both $z^m\det {\rm M}(z)$ and $\det [{\rm T}-zI_{2m}]$ 
are  polynomials in $z$ 
of degree $2m$ and share the same roots, which cannot be zero by 
(\ref{detT}). Therefore, the polynomials coincide up to a constant of 
proportionality, which is found by considering 
the limit case of large $z$: 
$\det{\rm M}(z)\approx (-1)^{nm}(-z)^m\, \det [B_1\cdots B_n]$. 
$\blacksquare $

\vskip0.5truecm
Before proceeding, let us show that in the special case of tridiagonal
matrices with corners ($m=1$), Lemma 1 with $z=1$ yields (\ref{trid0}). 

The factorization
\begin{equation}
\left(\begin{array}{cc}
-\frac{a_{k-1}}{b_{k-1}} & -\frac{c_{k-2}}{b_{k-1}} \\1 & 0 
\end{array}\right)=
\left(\begin{array}{cc} -\frac{1}{b_{k-1}} & 0 \\0 & 1 \end{array}\right)
\left(\begin{array}{cc}
a_{k-1} & -c_{k-2}b_{k-2} \\1 & 0 \end{array}\right)
\left(\begin{array}{cc} 1 & 0 \\ 0 & -\frac{1}{b_{k-2}}\end{array}\right)
\label{factorization}
\end{equation}
is introduced for all factors in 
the transfer matrix T and produces intermediate factors 
$\frac{1}{b_k}  I_2$ that commute, and allow us to simplify the determinant 
of the lemma:
\begin{eqnarray}
&&\det\left[
\left(\begin{array}{cc}-\frac{a_n}{b_n} & -\frac{c_{n-1}}{b_n} \\ 1 & 0 
\end{array}\right)\cdots
\left(\begin{array}{cc}-\frac{a_1}{b_1} & -\frac{c_0}{b_1} \\ 1 & 0 
\end{array}\right)- I_2\right]\nonumber \\
&&=\det\left[
\frac{(-1)^{n-1}}{b_1\cdots b_{n-1}}
\left(\begin{array}{cc} -\frac{1}{b_n}&0\\0&1\end{array}\right)
\left(\begin{array}{cc} a_n & -b_{n-1}c_{n-1} \\ 1 & 0 \end{array}\right)
\cdots\left(\begin{array}{cc} a_1 & -b_n c_0 \\ 1 & 0 \end{array}\right)
\left(\begin{array}{cc}1 & 0\\ 0 & -\frac{1}{b_n}\end{array}\right)- I_2\right]
\nonumber \\
&&=\frac{1}{b_1^2\cdots b_n^2}
\det\left[
\left(\begin{array}{cc} a_n & -b_{n-1}c_{n-1} \\ 1 & 0 \end{array}\right)
\cdots\left(\begin{array}{cc} a_1 & -b_n c_0 \\ 1 & 0 \end{array}\right)
-(-1)^n b_1\cdots b_n I_2\right]\nonumber \\
&&=\frac{z_1z_2}{b_1^{2}\cdots b_n^{2}}-(-1)^n \frac{z_1+z_2}{b_1\cdots b_n}+1
\nonumber\\
&&=-\frac{(-1)^n}{b_1\cdots b_n} 
[(z_1+z_2)-(-1)^n(b_1\cdots b_n + c_0\cdots c_{n-1})] \nonumber
\end{eqnarray}
$z_1$ and $z_2$ are the eigenvalues of the transfer matrix in 
(\ref{trid0}), whose trace is $z_1+z_2$ and whose determinant is 
$z_1z_2= (b_1\cdots b_n)(c_0\cdots c_{n-1})$. $\blacksquare $
\vskip0.5truecm

Multiplication of Lemma 1 by $\det{\rm T}^{-1}$ gives
a variant of it:  
\begin{displaymath}
\det {\rm M}(z)= (-1)^{nm}(-z)^m\, \det({\rm T}^{-1}-\frac{1}{z}) \,
\det[C_0\ldots C_{n-1}]
\end{displaymath}
Multiplication of Lemma 1 by the previous equation, with parameter $1/z$, 
gives another variant:
\begin{displaymath}
\det {\rm M}(z)\det {\rm M}(1/z)\,=\,
\det\left [{\rm T}+{\rm T}^{-1}-\left (z+\frac{1}{z}\right )\right ]
\, \det[B_1C_0\ldots B_nC_{n-1}]
\end{displaymath}
\vskip 0.5truecm

Instead of ${\rm M}(z)$, consider the matrix 
${\rm M}(z)-\lambda I_{nm}$ and the corresponding transfer matrix 
${\rm T} (\lambda)$ obtained by replacing the entries $A_i$ 
with $A_i-\lambda I_m$. Then Lemma 1 has a symmetric 
form, where the roles of eigenvalue and parameter exchange between
the matrices. For this reason it is called a {\em duality relation}.
\vskip0.5truecm
{\bf Theorem 1 (The Duality Relation)}\par
\begin{displaymath}
\det [\lambda I_{nm} - \,{\mathrm M}(z)]\,=\, (-z)^{-m}
\,\det [\,{\mathrm T}(\lambda )-zI_{2m}\, ] \,\det[B_1\cdots B_n]
\end{displaymath}
It shows that the parameter $z$, which enters in ${\rm M}(z)$ as a boundary 
term, is related to eigenvalues of the matrix ${\rm T}(\lambda )$ that 
connects the eigenvector of ${\rm M}(z)$ at the boundaries.

The duality relation was initially obtained and discussed for Hermitian 
block matrices\cite{Molinari97,Molinari98,Molinari02}. 
For $n=2$ it is due to Lee and Ioannopoulos\cite{Lee81}. 
Here I have shown that it holds for 
generic block-tridiagonal matrices, and the proof given is even simpler. 
The introduction of corner values $z$ and $1/z$ in Hermitian
tridiagonal matrices ($c_k=b_k^*$)
was proposed by Hatano and Nelson
\cite{Hatano96} in a model for vortex depinning in superconductors, as a tool
to link the decay of eigenvectors to the permanence of 
corresponding eigenvalues on the real axis. It has been a subject of intensive 
research\cite{Schnerb98,Feinberg99,Goldsheid98,Trefethen}.
The generalization to block matrices is interesting for the study of 
transport in discrete structures such as nanotubes or 
molecules\cite{Mahan99,Compernolle03,Yamada04}.

\section{Block tridiagonal matrix with no corners}
By a modification of the proof of the lemma, one obtains an identity for 
the determinant of block-tridiagonal matrices ${\rm M}^{(0)}$ with no corners 
($B_n=C_0=0$ in the matrix (\ref{mz})):
\begin{thm}
\begin{displaymath}
\det {\rm M}^{(0)} = (-1)^{nm} 
 \det [{\rm T}^{(0)}_{11}]\,\det[B_1\cdots B_{n-1}]
\end{displaymath}
where ${\rm T}^{(0)}_{11}$ is the upper left block of size $m\times m$ of 
the transfer matrix 
\begin{displaymath}
\mathrm{T}^{(0)}=\left[\begin{array}{cc}
-A_n & -C_{n-1} \\I_m & 0 \end{array}\right]
\left[\begin{array}{cc} 
-B_{n-1}^{-1}A_{n-1} & -B_{n-1}^{-1}C_{n-2} \\I_m & 0 \end{array}\right]
\ldots
\left[\begin{array}{cc} -B_1^{-1}A_1 & -B_1^{-1} \\I_m & 0 \end{array}
\right] 
\end{displaymath}
\end{thm}
Proof:
The linear system ${\rm M}^{(0)} \Psi =0$ can be translated into the following
equation, via the transfer matrix technique:
\begin{eqnarray}
\label{noc}
\left[\begin{array}{c} \psi_n \\ -C_{n-1}^{-1}A_n \psi_n \end{array}\right] 
&&\, = \,
\left[\begin{array}{cc}
-B_{n-1}^{-1}A_{n-1} & -B_{n-1}^{-1}C_{n-2} \\I_m & 0 
\end{array}\right]
\times\ldots \\
&&\times \left[\begin{array}{cc} 
-B_2^{-1}A_2 & -B_2^{-1}C_1 \\I_m & 0 
\end{array}\right] 
\left[\begin{array}{c} -B_1^{-1}A_1\psi_1 \\ \psi_1 \end{array}\right] 
\nonumber
\end{eqnarray}
Right multiplication by the nonsingular matrix
\begin{displaymath}
 \left[\begin{array}{cc} 
-A_n & -C_{n-1} \\I_m & 0 
\end{array}\right] 
\end{displaymath}
and rewriting the right-hand vector as the product
\begin{displaymath}
\left[\begin{array}{cc} 
-B_1^{-1}A_1 & -B_1^{-1} \\I_m & 0 \end{array}\right] 
\left[\begin{array}{c} \psi_1 \\ 0 \end{array}\right] 
\end{displaymath}
transform (\ref{noc}) into an equation for the transfer matrix 
${\rm T}^{(0)}$, that connects the boundary components with 
$\psi_{n+1}=0$ and $\psi_0=0$:
\begin{equation}
\label{eig2}
\left[\begin{array}{c} 0\\ \psi_n\end{array}\right] \,=\, {\rm T}^{(0)}
\left[\begin{array}{c}\psi_1\\ 0 \end{array}\right]
\end{equation}
Equation (\ref{eig2}) implies that $\det {\rm T}^{(0)}_{11}=0$, which is 
dual to $\det {\rm M}^{(0)}=0$. 
The implication translates into an identity by introducing the parameter 
$\lambda $ and comparing the polynomials 
$\det[\lambda I_{nm}-{\rm M}^{(0)}]$
and $\det \rm{T}^{(0)}(\lambda )$ (obtained by replacing blocks $A_i$ with
$A_i-\lambda I_m$). Since both are polynomials in 
$\lambda $ of degree $nm$ and with the same roots, they must be 
proportional. Their behaviour for large $\lambda $ fixes the constant. 
$\blacksquare $

For tridiagonal matrices ($m=1$) blocks are just scalars and, by means
of (\ref{factorization}), one shows Theorem 2 simplifies to (\ref{trid}).

The formula for the evaluation of $\det{\rm M}^{(0)}$
requires $n-1$ inversions $B_k^{-1}$, multiplication of $n$ matrices of 
size $2m\times 2m$, and the final evaluation of a determinant. 
Salkuyeh\cite{Salkuyeh06} proposed a different procedure for the 
evaluation of the same determinant:
\begin{displaymath}
\det {\rm M}^{(0)} = \prod_{k=1}^n \det\Lambda_k
\end{displaymath}
\begin{displaymath}
 \Lambda_k = A_k - C_{k-1}\Lambda_{k-1}^{-1} B_{k-1}, \qquad \Lambda_1=A_1
\end{displaymath}
It requires $n-1$ inversions of matrices of size $m\times m$, 
and the evaluation of their determinants. I show that the two procedures 
are related. 

The transfer matrix 
${\rm T}^{(0)}={\rm T}(n)$ is the product of $n$ matrices. Let $T(k)$ be the 
partial product of $k$ matrices. Then:
\begin{displaymath}
{\rm T}(k)\,=\,\left[\begin{array}{cc} 
-B_k^{-1}A_k & -B_k^{-1}C_{k-1}\\ I_m & 0
\end{array}\right] {\rm T}(k-1)
\end{displaymath}
This produces a two-term recurrence relation for blocks
\begin{displaymath}
{\rm T}(k)_{11}\,=\, -B_k^{-1}A_k {\rm T}(k-1)_{11} -B_k^{-1}C_{k-1}
{\rm T}(k-2)_{11}
\end{displaymath}
with $T(1)_{11}=-B_1^{-1}A_1$ and $T(0)_{11}=I_m$. The equations by Salkuyeh 
result for 
$\Lambda_k = -B_k{\rm T}(k)_{11}[{\rm T}(k-1)_{11}]^{-1} $.

%

%
\end{document}